\documentclass[aps,twocolumn,amsmath,amssymb,showpacs]{revtex4-1}
\usepackage{graphicx}
\usepackage{textcomp}
\usepackage{psfrag}
\begin{document}

\title{Dynamics of strain bifurcations in magnetostrictive  ribbon} 
\author{Ritupan Sarmah and G. Ananthakrishna}
\affiliation{Materials Research Centre, Indian Institute of Science, Bangalore 560012, India}

\begin{abstract}
We develop a coupled nonlinear oscillator model involving   magnetization and strain to explain several experimentally observed  dynamical features exhibited by forced magnetostrictive ribbon.  Here we show that the model recovers the observed period doubling route to chaos as function of the dc field for a fixed ac field and quasiperiodic route to chaos as a function of the ac field, keeping the dc field constant. The model also predicts induced and suppressed chaos  under the influence of an additional small-amplitude near resonant ac field. Our analysis  suggests rich dynamics in coupled order parameter systems like magnetomartensitic and magnetoelectric materials.
\end{abstract}


\maketitle

\section{Introduction}

The rich chaotic dynamics exhibited by sinusoidally driven nonlinear oscillators is ubiquitous to a large number of systems such as turbulence in fluid systems \cite{Gollub75}, chemical oscillators \cite{Swinney85}, and  cardiac tissues \cite{Glass83}. Manipulated properties of chaos, such as control of chaos \cite{LR03} realized in driven nonlinear oscillators, are also reported in several disciplines, for instance,  control of output of laser system \cite{Roy92}, enhanced performance of permanent magnet synchronous motor \cite{Liu04}, and control of cardiac arrhythmias \cite{Garfinkel92}. Natural to multistate weakly driven oscillators in the presence of ubiquitous  noise is the stochastic resonance \cite{Benzi81}, another generic noise induced cooperative phenomenon that enhances the signal to noise ratio. This phenomenon is also realized in several disciplines ranging from  physics ( e.g., superconducting quantum interference device magnetometers, optical and electronic devises) \cite{Gammaitoni98} to biology \cite{McDonnell09}. However, {\it it is rare to find such a broad spectrum of dynamical features in a single system}. Surprisingly, physical realization of several of these features was reported by Vohra {\it et~al.} \cite{Vohra91a,Vohra91b, Vohra94,Vohra95,Vohra93}. In their study of strain bifurcations in magnetostrictive ribbons subjected to the combined influence of sinusoidal (ac) and dc magnetic fields, the authors reported an extraordinarily large number of dynamical features such as (a)  quasiperioidic (QP) route to chaos when the amplitude of  the ac magnetic field $h_{ac}$ was increased in the presence of a  dc magnetic field $h_{dc}$, (b) period doubling (PD) route to chaos when  $h_{dc}$ was increased and the ac field was kept fixed \cite{Vohra91a,Vohra91b}, (c) a suppression and shift of period doubling bifurcation point \cite{Vohra91b} and  induced subcritical bifurcation under small-amplitude near-resonant conditions  \cite{Vohra94}, (d)  control of chaos, specifically, suppressed and induced chaos with the application of a near-resonant perturbation  to one of the subharmonics \cite{Vohra95}, and (e) stochastic resonance \cite{Vohra93}.  However, modeling such a rich dynamics exhibited by a {\it single} system in terms of the relevant strain and magnetic order parameters has remained  a challenge. 

Vohra {\it et~al.} emphasize that $\Delta E$ effect \cite{Savage86}, i.e., the reversible change in the Young's modulus with applied magnetic field, is not responsible for the reported dynamics as the samples used are unannealed ribbons where the $\Delta E$ effect is insignificant, unlike the earlier reported PD and QP routes to chaos in the annealed metallic glass samples \cite{Ditto89}.  Some  of these features have been explained using appropriate normal forms \cite{Bryant86} by appealing to the  universal nature  of the bifurcation phenomenon. Our purpose is to develop a model in terms of magnetic and strain order parameters. In this paper we focus on three different experimentally observed dynamical features and show that the model predicts (i) the period doubling and (ii) quasiperiodic routes to chaos and (iii) induced and suppressed chaos in the presence of  small-amplitude near-resonant perturbation.  The general nature of these equations suggest much richer dynamics in ferromagnetic martensite samples that posses even stronger elastic and magnetic nonlinearities  \cite{FMM} and also in magnetoelectric materials \cite{Flebig02}. The model also explains some old results on internal friction studies of martensites \cite{Suzuki80}.  

\section{The Model}

Experiments were  performed using a sinusoidal magnetic field $h_{ac}$ of frequency $\omega$ in the presence of a dc magnetic field $h_{dc}$. Our starting point is to write down the relevant free energies. In general, elastic free energy is a function of all components of the strain tensor. However, considering the fact that the samples are thin long ribbons ($5$cm long) fixed at one end with the displacement or strain being  monitored at the other end, it would be adequate to use a single strain order parameter $\epsilon$. Further, we  work in one dimension and use dimensionless  order parameters. We  construct a minimal one-dimensional model with the strain order parameter $\epsilon$  and  magnetic order parameter $m$. In our model the weak magnetic nonlinearity is taken to drive the highly nonlinear elastic degrees of freedom \cite{Vohra91a}. Then the total free energy has three contributions, namely, the strain free energy, magnetic free energy, and magneto-elastic free energy.  Thus the total free energy is $F_T=F_{\epsilon} + F_m + F_{m\epsilon}$.

In one dimension the elastic free energy is given by 
\begin{equation}
F_{\epsilon}=\int dy'\left(f_{local}(\epsilon(y'))+\frac{1}{2}(\frac{\partial \epsilon}{\partial y'})^2 \right), 
\end{equation}
where the strain variable is $\epsilon = \frac{\partial u(y)}{\partial y}$, $f_{local}$ is the Landau free energy density, and the second term is the gradient free energy. Guided by the normal forms used \cite{Vohra91a,Vohra91b, Vohra93,Vohra94,Vohra95}, we use  a sixth-order polynomial for the elastic free energy $f_{local} = \frac{\theta}{2} \epsilon^2 -\frac{\beta}{2} \epsilon^4 + \frac{\Delta}{6} \epsilon^6$. Here $\beta$ is a positive constant, but $\theta$ can take on positive or negative values. To simplify further, we use $\Delta=1$. The minima of the free energy are located at $\epsilon = \pm \epsilon_s$ with $\epsilon_s =[(\beta + \sqrt{\beta^2 -\theta})]^{1/2}$.  The parameters $\beta$ (of the order of unity) and $\theta$ are sufficient to control the minima.  When $\theta \ge 3\beta^2/4$, $\epsilon=0$ is the true minima identified with the high-temperature phase. At $\theta =3\beta^2/4$, we have a first order transition with the free energy vanishing at $\epsilon^2 =\pm 3\beta/2$ and $\epsilon=0$.  For $\theta$ negative, $\epsilon=0$ state is unstable. This behavior can be parameterized by using $\theta =  \frac{T- T_s}{T_f - T_s} =\tau_{\epsilon}$ where $T_f$ and $T_s$ are the first- and second order transition temperatures, respectively. 

The magnetic free energy is given by
\begin{eqnarray}
\nonumber
F_m = \int dy' \big[ f_{mag}+\frac{1}{2}(\frac {\partial m}{\partial y'})^2 - m(h_{dc} + h_{ac}\sin \omega t)\big],
\end{eqnarray}
where $f_{mag} = \tau_c\frac{m^2}{2} + \frac{m^4}{4}$ with $\tau_c = \frac{T- T_c}{T_c}$, where $T_c$ is the Curie temperature of the sample. The nature of magneto-elastic free energy is not known, particularly since the ribbon is a metallic glass sample with a glassy structure. However, coupling terms either preserve or break the invariance $\epsilon \rightarrow -\epsilon$ and $m\rightarrow -m$ in the elastic and magnetic free energies, respectively. In the absence of any information,  we model it  as a weighted sum of symmetry-preserving and -breaking terms given by
\begin{equation}
F_{m\epsilon}=- \frac{\xi}{2}\int dy'[(1-p)\epsilon(y')m(y') + p\epsilon^2(y')m^2(y')]
\end{equation}
where $\xi$ is magnetoelastic coupling coefficient and $ 0 \le p \le 1$ is an adjustable weight factor. Other types of gradient coupling between $\epsilon$ and $m$ are ignored to keep the model simple.  We use the Rayleigh dissipation function \cite{Landau} to represent the damping of oscillating ribbon when the applied field is removed. This is given by
\begin{equation}
F_{diss}=\frac{\gamma}{2}\int\!\!\left(\frac{\partial \epsilon}{\partial t}\right)^2 dy'.
\end{equation}

The kinetic energy  is given  by $T = \frac{1}{2} \int \left[\frac{\partial u(y')}{\partial t}\right]^2 dy'$, where $u(y,t)$ is the displacement variable. Then, using the Lagrangian $L= T - F_T$ and the Lagrange equations of motion $\frac{d}{dt}\left(\frac{\delta L}{\delta \dot{u}(y)}\right)-\frac{\delta L}{\delta{u}(y)} =  {-\frac{\delta F_{diss}}{\delta \dot{u}(y)}}$,
we get
\begin{eqnarray}
\frac{\partial^2 \epsilon(y)}{\partial t^2} &=& \frac{\partial^2}{\partial y^2} \Big[ \tau_{\epsilon} \epsilon(y)-2 \beta \epsilon^3(y)+ \epsilon^5(y)- \frac{\partial^2}{\partial y^2} \epsilon(y) \nonumber \\
 &+&\gamma\frac{\partial \epsilon(y)}{\partial t} -\xi \big(p \epsilon(y) m^2(y)+ \frac{(1-p)}{2} m(y)\big)\Big]. 
\label{S}
\end{eqnarray}

Using the equation of motion for the magnetic order parameter given by $\frac{\partial m}{\partial t} = - \Gamma \frac{\delta F_T}{\delta m}$,  we get 
\begin{eqnarray}
\label{M}
\frac{\partial m(y)}{\partial t}&=&-\Gamma \Big[\tau_c m(y) + m^3(y) -  \frac {\partial^2 m(y)}{\partial y^2}- h_{dc}  \nonumber \\
&-& h_{ac} \sin \omega t - \xi\Big( p \epsilon^2(y) m(y)+ \frac{(1-p)}{2}\epsilon(y)\Big)\Big],
\end{eqnarray}
where $\Gamma$ is a relevant time scale. Note that this is not a pure relaxational dynamics as $m(y)$ is subject to sinusoidal and dc fields. Indeed, $\epsilon(y)$ is being driven through $m(y)$, which is subject to the sinusoidal field. [Note also the mutual coupling between $m(y)$ and $\epsilon(y)$.] For a sample that is clamped at one end, the above equations can be further simplified by noting that the strain at the free end is small even though it is in the anharmonic regime. Thus we assume that  only the dominant mode of vibration is supported. We use $\epsilon(y,t) = A(t) \sin k y$ and $m(y,t) = B(t) \sin ky$, where $k= \pi/2L$, with $L$ representing the length of the ribbon. Further, we note that the dc field induces a finite strain value, which can be determined by equating the strain energy with the magnetic energy. The application of the ac magnetic field induces oscillations around this strain value. Thus we use  the dominant mode to represent deviation from the equilibrium value of the dc field, i.e., $A = a-a_{dc}$. Here, $a$ is the strain amplitude in the presence of both magnetic fields and $a_{dc}$ is that when only the dc field is imposed. Transforming  Eqs.(\ref{S},\ref{M}) into rescaled variables ($\tau=kt, \Omega= \omega/k, \Gamma'=\Gamma/k$) and using only the dominant mode, we get 
\begin{eqnarray}
\nonumber
\ddot{A}(\tau)&=&-\Big[\tau_{\epsilon} A(\tau)+3 \beta A^3(\tau)-\frac{5}{8}A^5(\tau)+k^2A(\tau) \\
\label{A}
&&+k\gamma\dot{A}(\tau)+\frac{\xi}{2} \Big( 3 pA(\tau)B^2(\tau) - (1-p) B \big) \Big],\\
\nonumber
\dot{B}(\tau)&=&-\Gamma' \Big[\tau_cB(\tau)+\frac{B^3(\tau)}{2}+k^2 B(\tau) -\frac{\xi}{2} \Big( (1-p)A\\
&& +  p A^2(\tau)B(\tau) \Big)- h_{ac} \sin \Omega \tau - h_{dc} \Big].
\label{B}
\end{eqnarray}
The overdot now refers to the redefined time derivative.  Equations (\ref{A}) and (\ref{B}) constitute a coupled set of nonlinear nonautonomous ordinary differential equations with several parameters. We shall use $\tau_{\epsilon}, \tau_c, \gamma,\Gamma', \Omega$, and $\xi$ as free parameters.  In addition, both $h_{ac}$ and $h_{dc}$ are the experimental drive parameters. Apart from this, we have no knowledge of the values of the parameters. 

\section{Results}

\begin{figure}[t]
\vbox{
\includegraphics[height=3.0cm,width=7.0cm]{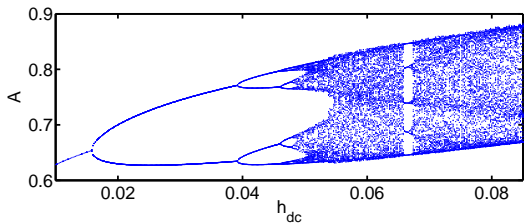}
}
\caption{(Color online) Period-doubling bifurcation as a function of $h_{dc}$ for $\tau_c = -0.2, \tau_{\epsilon}=-1.0,\Omega=1,\xi=0.6, \gamma= 1.592, \Gamma'=0.09, h_{ac}=10.5$, and $p=0.32$. 
}
\label{PD-Bif}
\end{figure}

The dynamics of the above system of equations is quite complicated due to the presence of several parameters.  Clearly, interesting dynamics can be expected when $\tau_{\epsilon}$ is in the range where three or two  minima exist and when $\tau_c$ is negative. Thus we first map-out the region of the parameter space where interesting dynamics is seen.  

Consider the period-doubling route as a function of $h_{dc}$ for a fixed  $h_{ac}$. Here, we keep $\tau_{\epsilon}=-1$ so that $\epsilon=0$ state is unstable, although interesting dynamics is also observed for $\tau_{\epsilon} < 3\beta^2/4$ as well.  Other parameters are fixed at $\tau_c = -0.2, \Omega = 1, \gamma= 1.592, \Gamma'= 0.09, L =5, \xi=0.6$, and $p=0.32$. 

\begin{figure}
\vbox{
\includegraphics[height=3.75cm,width=7.5cm]{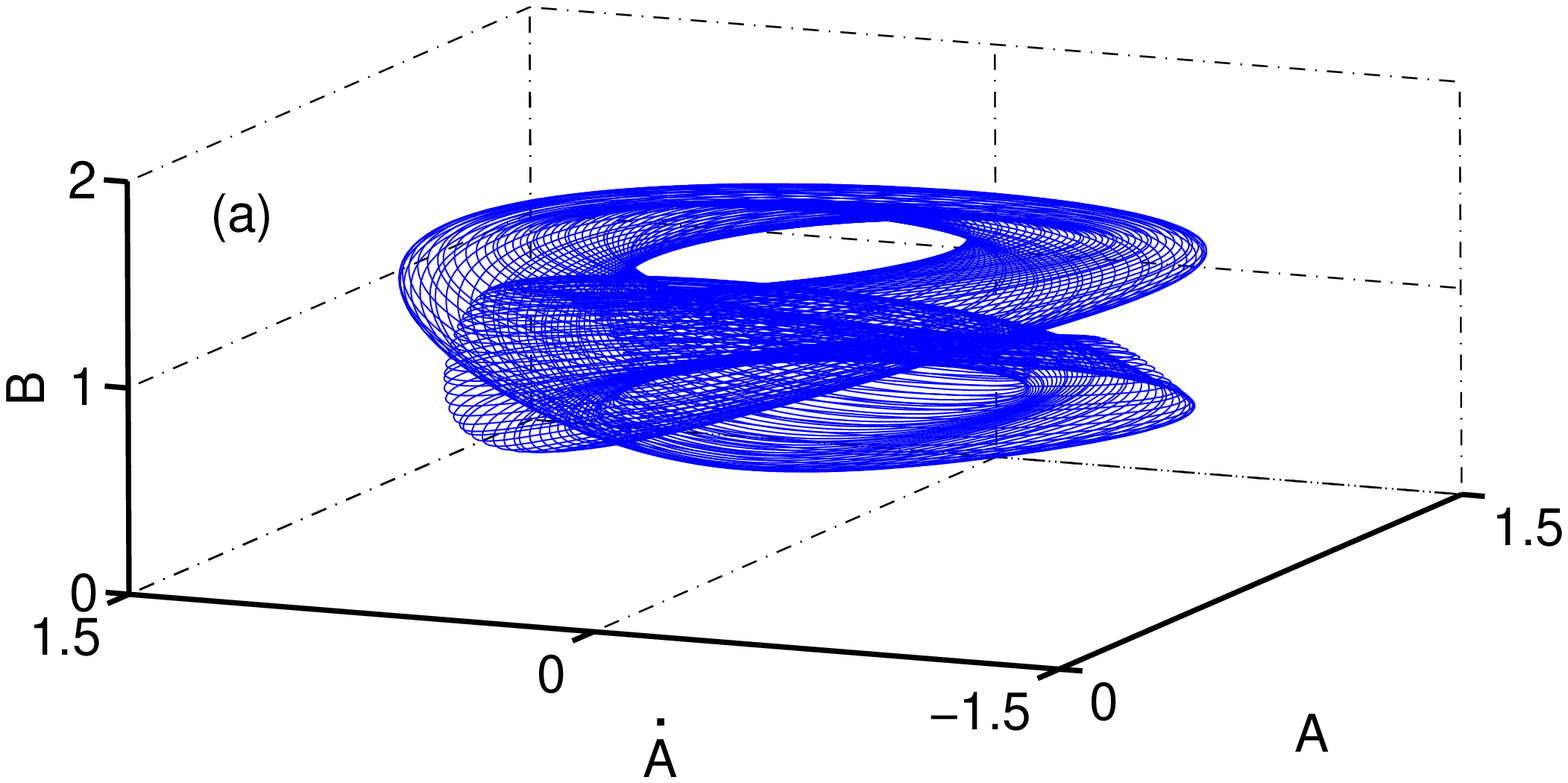}
\includegraphics[height=2.5cm,width=8.4cm]{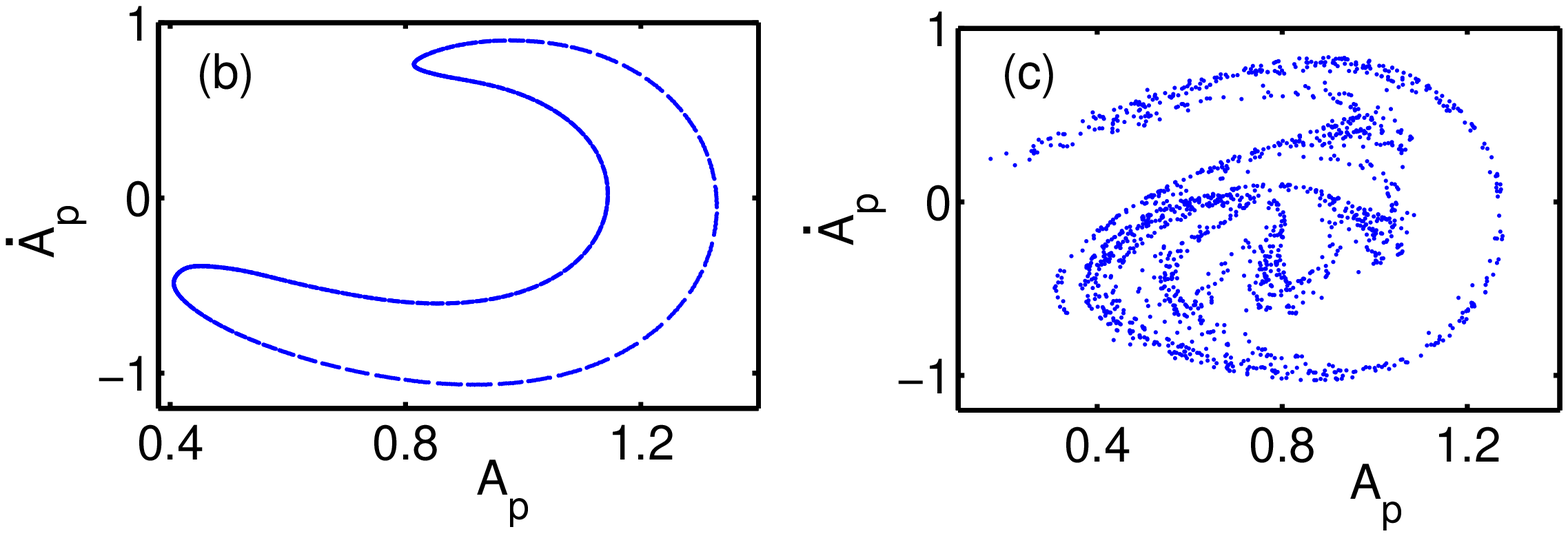} 
}
\caption{(Color online)(a) Quasiperiodic orbit in the $(\dot A, A, B)$ space for  $h_{ac}=4.8$ keeping $\tau_c = -0.2, \tau_{\epsilon}=-1.0, \Omega =0.97, \gamma =0.0421  ,\xi=3.2, \Gamma'=0.09, h_{dc}= 0.04$, and $p=0.32$. (b) The corresponding Poincar\'e map in the $(A_p, \dot A_p)$ plane. (c) Poincar\'e map for a chaotic orbit for $h_{ac} = 7.4$.}
\label{QP-Route}
\end{figure}

The application of the dc field in the presence of ac field has two effects. First, the vibrations are centered around the strain induced by the dc field. This has already been included in the definition of $A$. Second,  the free energy for the magnetic  order parameter is tilted to one side. In addition, the shape of the free energy is also affected due to the presence of nonlinear  magnetoelastic coupling. In particular,  note that apart from the  additive $B$ term in Eq. (\ref{A}), $B^2$ appears multiplicatively with $A$ (equivalent to parametric forcing). This affects the shape of the effective elastic free energy. Thus we should expect $\beta$ to be a function of $h_{dc}$.  To understand this,   we first note that in the absence of $h_{dc}$, increasing the amplitude of the ac field gives period-doubling bifurcation, as expected from studies on Duffing-like oscillators \cite{Musielak05}.  Increasing $\beta$ also  leads to period-doubling sequence keeping $h_{dc}=0$, but keeping $h_{ac}$ at a value where we find period one. In contrast, keeping $h_{ac}$  at a values where chaos is seen (with $\beta=1$) and  increasing $h_{dc}$ reverses the period doubling sequence, reflecting the compensating relationship between $\beta$ and $h_{dc}$. This can be made quantitative by  keeping $h_{ac} = 9.2$ (a value  corresponding to a period-four orbit) and increasing both $h_{dc}$ and $\beta$ such that we always see the period-four orbit. This gives a relation $\beta= 1+ 10 h_{dc}$. However, while $\beta$ is linear in $h_{dc}$, the prefactor depends on the value of $h_{ac}$ used.

For further calculations, we use this parameterized form of $\beta$ in Eqs.(\ref{A},\ref{B}). Then, increasing $h_{dc}$ and keeping $h_{ac} = 10.5$  leads to period-doubling  route to chaos.  All quantitative measures are obtained after discarding the first $1.5 \times 10^5$ time steps. A plot of PD sequence is shown in Fig. \ref{PD-Bif}. The PD route is seen for a range of  parameter values around those used for Fig. \ref{PD-Bif}.

\begin{figure}
\vbox{
\includegraphics[height=3.5cm,width=7.5cm]{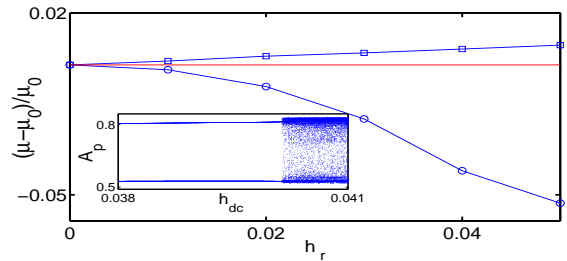}
}
\caption{(Color online) Suppressed and induced chaos for $\delta =10^{-6}(\square)$  and $\delta=10^{-2}(\circ)$ respectively as a function of $h_r$. The inset shows the  bifurcation diagram for $\tau_c = -0.2, \tau_{\epsilon}=-1.0,\xi=1.0, \Omega =1, \gamma= 1.592,\Gamma'=0.09, h_{ac}= 10.5$, and $p=0.32$.}
\label{Shift-Chaos}
\end{figure}

Now consider quasiperiodic route to chaos. For this case we choose the driving frequency to be less than that for the period-doubling route ($\Omega =1$). For the results reported here $\Omega =0.97$, but the similar results are obtained for a range of values of $\Omega$ less than unity.  We retain most values of the parameters as in the PD sequence, but use higher coupling $\xi =3.2$ and $\gamma= 0.0421$. In this case, keeping $h_{dc} =0.04$ fixed, we sweep $h_{ac}$ from $2$ onwards. We find the first few harmonic frequencies (as found in turbulence of rotating fluid; see Ref. \cite{Gollub75}) in the region $h_{ac} = 2.5$ to $3.8$,  beyond which quasiperiodicty is seen until $5.1$ with the emergence of two incommensurate frequencies.  A plot of the torus in the $(A, \dot A, B)$ space is shown in Fig. \ref{QP-Route}(a) for $h_{ac}=4.8$. The corresponding Poincar\'e map in the $(A_p,\dot A_p)$ plane is shown in Fig. \ref{QP-Route}(b). (The largest Lyapunov exponent vanishes in the regions of quasiperiodicity.) We find three  regions  of chaos interrupted by windows of quasiperiodicity in the interval $h_{ac}= 2.0$ to $7.5$. The first region of chaos is seen between $h_{ac}= 5.45$ to $5.53$ preceded by a phase locked region. This is followed by a transition to quasiperiodic regime ($h_{dc} = 5.53 -5.64$) followed by an abrupt transition to  chaos ($h_{dc} = 5.64 - 5.73$). Another region of chaos is seen for $h_{ac}= 7.3- 8.0$. The Poincar\'e map for the chaotic region ($h_{ac}=7.4$ ) is shown in Fig. \ref{QP-Route}(c).  The QP route to chaos is observed for a range of values of parameters around the values used for Fig. \ref{QP-Route}. 

We next examine the possibility of induced and suppressed chaos when the system is subjected to small-amplitude near-resonant perturbation of the form $ h(\tau) = h_r \sin \Omega_r \tau$, with $\Omega_r$ close to the first  or the second subharmonic. For illustration we use $\Omega_r = \Omega/2 - \delta$, keeping  $\delta$ small.  To do this we first locate a direct transition from period-two cycle to chaos as a function of $h_{dc}$, keeping perturbing field $h_r=0$. We find that this transition occurs when $h_{dc} = 0.04014$ (except for $\xi=1.0$, all other parameters values are the same as for PD route). A bifurcation diagram for  the strain amplitude is shown in the inset of Fig. \ref{Shift-Chaos}. Keeping $h_{dc}=0.04014$, we apply the perturbing signal $h_r \sin \Omega_r \tau$ with $\delta = 10^{-6}$. Even for small $h_r$ we find that the onset of chaos is delayed. Further, the magnitude of the shift, though small, increases linearly with $h_r$. Identifying $\mu_0$  with the onset of chaos ($h_{dc} =0.04014$) when $h_r =0$  and $\mu$ with that for finite $h_r$,  the normalized shift ($\square$) $\frac{\mu-\mu_0}{\mu_0}$ for the onset of chaos (obtained by computing  the Lyapunov exponent) as function of $h_r$  is shown in Fig. \ref{Shift-Chaos}.  We have also studied the influence of $\delta$ on the nature of the dynamics. For instance, for $\delta = 10^{-2}$, we find induced chaos. The magnitude of the shift in the onset of chaos ($\circ$), which is significantly higher than that for the suppressed chaos, is shown in Fig. \ref{Shift-Chaos}.

\section{Summary and Conclusions}

In summary, we have developed a coupled oscillator model where the weak magnetic nonlinearity drives the highly nonlinear strain order parameter.  The model explains several experimental results on the dynamics of the magnetostrictive ribbon such as  (a) the period-doubling bifurcation as a function of $h_{dc}$ and keeping $h_{ac}$ fixed, (b) the quasiperiodic route to chaos as a function of $h_{ac}$ for a fixed dc field, and (c) induced and suppressed chaos under the influence of the resonant perturbing field  $h_r$ \cite{Vohra91a,Vohra91b, Vohra93,Vohra94,Vohra95}. However, the suppressed chaos seen here is for small $\delta$ while induced chaos is seen for large $\delta$, which is the opposite of what has been reported. This may be attributed to parametric type of forcing in our equations. Note that the method of control of chaos is different from traditional methods \cite{LR03}. The model demonstrates induced and suppressed chaos in the presence of near-resonant perturbation. The model exhibits rich dynamics including transient chaos. For instance, for  the symmetry restoring crisis \ref{Ishii86}, we find that the mean time $<\tau>$ spent in the precrisis attractor scales as $<\tau> \sim (h_{ac} - h_{ac}^c)^ {- \eta}$, where $h_{ac}^c$ is the critical value, with $\eta \sim 1.06 $.

The approach is clearly applicable to any coupled  order parameter system \cite{FMM,Flebig02}  such as magnetomartensites \cite{FMM}, which exhibits high elastic strain and  magnetization, and ferroelectromagnetic materials, which exhibit magnetization and electric polarization \cite{Flebig02}. Such systems are also described by polynomial forms of free energy, as used in the model. Our analysis suggests rich dynamics if experiments are performed in a similar geometry on these materials. This should encourage dynamic experiments in these  materials where experiments are traditionally carried out in quasistatic conditions.

The model has been adopted  to explain  some old unexplained results \cite{Ritupan11} in the internal friction experiments on samples of nonmagnetic martensite samples of $Cu_{82.9}Al_{14.1}Ni_3$ reported \cite{Suzuki79,Suzuki80}. These experiments have also been carried out in a similar geometry (but transverse drive) near  martensite transformation temperature $T_m$, where the elastic nonlinearities are significant.   Our equations automatically describe the twinned structure in one dimension when the magnetic free-energy contribution is dropped \cite{Bales}. The model recovers all results \cite{Ritupan11} including the  period four cycle (which the authors  even fail to recognize; see Fig. 2 of \cite{Suzuki80}).

These equations in their present form describe the dynamics of magnetomartensites as well \cite{FMM}.  These alloys  undergo a first-order martensitic transformation on cooling below $T=T_m$ and are also ferromagnetic below the Curie temperature $T_c$.  Below $T_m$ and $T_c$ they have high degree of elastic and magnetic nonlinearities. Indeed, the application of a small amount of magnetic field induces large strains by the rearrangement of martensite variants \cite{FMM}. Similarly, stress also affects magnetization, thereby suggesting strong magnetoelastic coupling. Thus a rich dynamics is predicted in magnetomartensites if experiments are performed in a similar geometry. It would be interesting to verify this prediction. 

Finally, since the application of magnetic field can induce large strains (as high as $\sim 10\%$), magnetomartensites are better suited for actuator applications than even the best magnetostrictive materials (such as Terfenol-D, with a strain of $\sim 0.2\%$ ). Further, these materials should provide more accurate control of the measurement of dynamic strains compared to magnetostrictive materials \cite{Vohra94}.  

\begin{acknowledgments}
G. A acknowledges financial support from Indian National Science Academy  and Board of Research in Nuclear Sciences (BRNS) Grant No. 2007/36/62-BRNS/2564. RS acknowledges Council of Scientific Industrial
Research, India for financial support.
\end{acknowledgments}


\begin{thebibliography}{99}
\bibitem{Gollub75} J. P. Gollub and H. J. Swinney, Phys. Rev. Lett. {\bf 35}, 927 (1975).
\bibitem{Swinney85} J. Maselko and H. L. Swinney, Phys. Scr. {\bf T9}, 35 (1985).
\bibitem{Glass83} L. Glass {\it et al.}, Physica D {\bf 7}, 89 (1983).
\bibitem{LR03} M. Lakshmanan and S. Rajasekar, {\it Nonlinear Dynamics, Integrability, Chaos and Pattern Formation} (Springer-Verlag, Heidelberg, 2003).

\bibitem{Roy92} R. Roy, T. W. Murphy, T. D. Maier, Z. Gills, and E. R. Hunt, Phys. Rev. Lett. {\bf 68}, 1259 (1992).
\bibitem{Liu04} D. Liu, H. P. Ren and X.Y. Liu, Proceedings of the 2004 International Symposium on Circuits and Systems, 2004, ISCAS '04 [IEEE Circuits Syst. {\bf 4}, 732 (2004)].
\bibitem{Garfinkel92} A. Garfinkel {\it et~al.}, Science {\bf 257}, 1230 (1992).
\bibitem{Benzi81} R. Benzi, A. Sutera and A. Vulpiani, J. Phys. A {\bf 14}, L453 (1981).
\bibitem{Gammaitoni98} L. Gammaitoni {\it et~al.}, Rev. Mod. Phys. {\bf 70}, 223 (1998).
\bibitem{McDonnell09} M. D. McDonnell and D. Abbott, PLoS Comput. Biol. {\bf 5}, e1000348 (2009).
\bibitem{Vohra91a} S. T. Vohra {\it et al.},  J. Appl. Phys. {\bf 69}, 5736 (1991).
\bibitem{Vohra91b} S. T. Vohra, F. Bucholtz, K. P. Koo, and D. M. Dagenais, Phys. Rev. Lett. {\bf 66}, 212 (1991).
\bibitem{Vohra94} S. T. Vohra, L. Fabiny, and  K. Wiesenfeld, Phys. Rev. Lett. {\bf 72}, 1333 (1994).
\bibitem{Vohra95} S. T. Vohra, L. Fabiny, and  F. Bucholtz, Phys. Rev. Lett. {\bf 75}, 65 (1995).
\bibitem{Vohra93} S. T. Vohra and  F. Bucholtz, Phys. Rev. Lett. {\bf 70}, 1425 (1993); S. T. Vohra and  L. Fabiny, Phys. Rev. E {\bf 50}, R2391 (1994).

\bibitem{Savage86} H. T. Savage and C. Adler, J. Magn. Magn. Mater. {\bf 58}, 320 (1986).

\bibitem{Ditto89} W. L. Ditto {\it et al.}, Phys. Rev. Lett. {\bf 63}, 923 (1989).
\bibitem{Bryant86} P. Bryant and K. Wiesenfeld,  Phys. Rev. A {\bf 33}, 2525 (1986). 
\bibitem{FMM} See, for example, R. Kainuma {\it et al.}, Nature {\bf 439}, 957 (2006), and the references therein.
\bibitem{Flebig02} M. Fiebig {\it et al.}, Nature (London) {\bf 419}, 818 (2002); W. Eerenstein, N. D. Mathur, and J. F. Scott, {\it ibid} {\bf 442}, 759 (2006).
\bibitem{Suzuki80} M. Wuttig and T. Suzuki, Scr. Metall. {\bf 14}, 229 (1980).

\bibitem{Landau} L. D. Landau and E. M. Lifschitz, {\it Theory of Elasticity}, 3rd ed. (Pergamon, Oxford, 1986).

\bibitem{Musielak05} D. E. Musielak, Z. E. Musielak, and J. W. Benner, Chaos Solitons Fractals {\bf 24}, 907 (2005).

\bibitem{Ishii86} H. Ishii, H. Fujisaka and M. Inoue, Phys. Lett., A{\bf 116}, 257 (1986).

\bibitem{Ritupan11} R. Sarmah and G. Ananthakrishna (unpublished).

\bibitem{Suzuki79} M. Wuttig and T. Suzuki, Acta Metall. {\bf 27}, 755 (1979).
\bibitem{Bales} G. S. Bales and R. J. Gooding, Phys. Rev. Lett. {\bf 67}, 3412 (1991). 

\end{thebibliography}
\end{document}